\documentclass[journal=jacsat,manuscript=article]{achemso}

\usepackage[version=3]{mhchem} 
\usepackage[utf8]{inputenc}
\usepackage[T1]{fontenc}
\usepackage{lmodern}
\usepackage{amsmath}
\usepackage{soul}
\usepackage{amsfonts}
\usepackage{amsthm}
\usepackage{amssymb}
\usepackage{newunicodechar}
\newunicodechar{ℝ}{\mathbb{R}}
\usepackage[sort&compress,numbers,super]{natbib}
\usepackage{achemso} 

\author{Bibekananda Das$^\mathsection$}
\affiliation[Unknown University]
{School of Physical Sciences, National Institute of Science Education and Research (NISER)
Bhubaneswar, An OCC of Homi Bhabha National Institute, Jatni 752050, Odisha, India}
\author{Tapas Senapati$^\mathsection$}
\affiliation[Unknown University]
{School of Physical Sciences, National Institute of Science Education and Research (NISER)
Bhubaneswar, An OCC of Homi Bhabha National Institute, Jatni 752050, Odisha, India}
\author{Malaya K. Sahoo}
\affiliation[Unknown University]
{School of Chemical Sciences, National Institute of Science Education and Research (NISER)
Bhubaneswar, An OCC of Homi Bhabha National Institute, Jatni 752050, Odisha, India}

\author{Jogendra N. Behera}
\affiliation[Unknown University]
{School of Chemical Sciences, National Institute of Science Education and Research (NISER)
Bhubaneswar, An OCC of Homi Bhabha National Institute, Jatni 752050, Odisha, India}
\alsoaffiliation{Center for Interdisciplinary Sciences (CIS),
National Institute of Science Education and Research (NISER),
An OCC of Homi Bhabha National Institute (HBNI),
Jatni, 752050, Odisha, India}
\email{jnbehera@niser.ac.in}
\author{Kartik Senapati}
\affiliation[Unknown University]
{School of Physical Sciences, National Institute of Science Education and Research (NISER)
Bhubaneswar, An OCC of Homi Bhabha National Institute, Jatni 752050, Odisha, India}
\alsoaffiliation{Center for Interdisciplinary Sciences (CIS),
National Institute of Science Education and Research (NISER),
An OCC of Homi Bhabha National Institute (HBNI),
Jatni, 752050, Odisha, India}
\email{kartik@niser.ac.in}

\title[An \textsf{achemso} demo]
  {Sensing Magnetic Flux of Langmuir-Blodgett Films of a Molecular Magnetic System using Superconducting Films and Nano-SQUID Devices}

\abbreviations{}

\begin{document}
\def\thefootnote{§}\footnotetext{These authors contributed equally to this work}







\begin{abstract}
We report a study on the response of superconducitng micro-tracks and quantum interference devices (SQUIDs) to a proximal single molecule magnet (SMM) film. As a test case, Langmuir-Blodgett $Mn_{12}$-ac SMM films have been grown on 2 $\mu$m wide Nb tracks and Nb nano-SQUIDs to observe the proximity effect of magnetic moment and magnetization tunneling, respectively. The superconducting critical temperature of thin Nb tracks (thinner than the coherence length of Nb) were found to decrease by the magnetic moment of $Mn_{12}$-ac SMM. Following the thermally activated flux flow (TAFF) model, we found an increase in the vortex unbinding energy of the SMM coated Nb tracks, near critical temperature. More importantly, the random alignment of moments of the $Mn_{12}$-ac molecules at low fields seemed to have the enhancing effect on vortex unbinding energy rather than the saturated state of $Mn_{12}$-ac molecules at high fields. In the fully  superconducting state, on the other hand, the vortex pinning effects were found to be more effective in the saturated state of the $Mn_{12}$-ac molecules, as seen from magnetoresistance and field dependent critical current measurements. In a separate experiment, a Langmuir-Blodgett film of SMM was grown on a nano-SQUID to look for local changes in magnetization arising from magnetization tunnelling phenomenon in SMMs. Upon magnetizing the SMM (deposited on SQUIDs) at 2 K  along the plane of the film and allowing it to relax, we found occasional jumps in the underlying SQUID voltage, unlike bare nano-SQUIDs, which did not show any such jumps over several hours. Therefore, we believe that the jumps in the SQUID voltage are the signatures of random tunneling of magnetization in the SMM layer.  

\end{abstract}

\section{Introduction}
In recent years, intense research has been carried out on development of single molecule magnetic (SMM) systems  because of their potential application in the field of molecular spintronics and quantum computing\cite{bogani2008molecular, PhysRevB.99.245404, leuenberger2001quantum, gaita2019molecular}. Among various SMMs 
$Mn_{12}$-acetate ($Mn_{12}$-ac) is the most widely studied system which was first synthesized by Lis et al. \cite{lis1980preparation}. The molecule consists of four $Mn^{4+}$ and eight $Mn^{3+}$ ions, and these Mn ions are coupled by superexchange interaction through oxygen bridges. These magnetically anisotropic high spin $(S=10)$ molecules show slow magnetic-relaxation because of strong anisotropic barrier and exhibit quantum tunneling of magnetization (QTM) behavior through this barrier below its blocking temperature ($T_B\sim$ 4 K)\cite{sessoli1993magnetic, thomas1996macroscopic}. 
Several techniques have been employed to prepare SMM thin films such as dip and dry\cite{10.1063/1.1808492, 10.1063/1.2800829}, drop and dry\cite{SEO200631}, and laser ablation \cite{MEANS2004215}methods towards the goal of device applications.  However, Langmuir-Blodgett (LB) technique is suitable for controlled layer growth \cite{oliveira2022past, ariga2020don, clemente1998langmuir}, which is required for devide applications. Clemente-LeoÂn \textit{et al.} have reported LB film of $Mn_{12}$-ac by mixing it with behenic acid (BA) lipid (with chemical formula $C_{21}H_{43}COOH$)\cite{clemente1998langmuir}. 

For spintronics applications, it is also important to study the response of hybrid systems consisting SMMs and other spintronics materials. Superconductor-SMM hybrid system is interesting because though ferromagnetism and superconductivity are two antagonistic effects, they find interesting ways to coexist\cite{bhatia2021nanoscale} in such structures. Serrano \textit{et al.} have grown a submonolayer of $Fe_4$ SMMs on Pb(111) and showed that the superconducting transition to the condensate state switches the SMM from a blocked magnetization state to a resonant quantum tunnelling regime\cite{serrano2020quantum}. Yu-Shiba-Rusinov bound state has been noticed in Co-porphyrin molecule with the Cooper pair condensate of Pb(111)\cite{schulte2023changing}. Inverse spin-valve effect and a significant alterations of the magneto-transport below and above 50 K by the SMM have been reported in ferromagnetic $La_{0.7}Sr_{0.3}MnO_3$ \cite{cucinotta2017tuning}. A giant magnetoresistance up to 1000\% and supramolecular spin-valve effect have been reported in case SMM on CNT \cite{krainov2017giant}. 
Earlier, a Josephson junction has been used as a radiation detector of electromagnetic radiation emitted during magnetization avalanches of an approximate cube of side 3 mm of 20 mg of $Mn_{12}$-ac placed near to it \cite{PhysRevB.70.140403}. Bellido et al. deposited $Mn_{12}$-benzoate inside a sensitive $\mu$-SQUID to measure its ac susceptibility\cite{C3NR02359A}. Due to their SMM transfer technique and the geometry of rf-SQUID, they were able to study a large collection of SMM particles. 


Fundamentally superconductors respond to magnetic flux either via breaking of Cooper pairs or via flux quantization. The superconducting properties of a Superconductor-SMM hybrid system, therefore, can be expected to show a flux-sensitive response. In this work, we have tried to establish the feasibility of detecting the flux response of a very low-density Langmuir-Blodgett SMM film using superconducting films and devices. While critical parameters of thin superconducting films can provide indirect and average information on the proximity effect of SMMs, nanoscale quantum interference devices can probe local effects at the single-molecule level. In this context, here we have studied Nb micro-tracks coated with LB film of $Mn_{12}$-ac molecular magnets embedded in the Behenic acid matrix. By comparing the transport data of SMM coated Nb tracks with bare Nb tracks we show that vortex pinning strength at low temperatures and vortex unbinding energy in the superconductor to normal transition regime enhances in presence of the proximal SMM laper. For capturing signatures of local magnetization tunnelling effects in SMM we have used Nb nano-SQUIDs. The sensitivity of a SQUID device is directly proportional to the loop inductance and capacitance. By lowering the capacitance and size of the junction or/and decreasing the loop diameter, sensitive nano-SQUIDs can be fabricated. While the small size improves flux noise resolution, and also functional up to very high field. The constriction-based (Dayem bridge) Nb- nano SQUID is a good choice for the purpose of the current study. By studying the time response of the voltage across a $Mn_{12}$-ac coated Nb nano-SQUID we show a possible signature of magnetization tunnelling as a voltage jump corresponding to a sudden change in the local moment in the vicinity of the SQUID loop. 

\section{Experimental Section}
\subsection{Preparation of $Mn_{12}$-ac powder}
The $Mn_{12}$-ac powder was prepared according to a previously reported method\cite{VERMA201776}. The $Mn_{12}$-ac was prepared by mixing 4.0 g of manganese acetate tetrahydrate ($Mn{(CH_3COO)}_2$ · $4H_2O$) with 40 mL of 60\% acetic acid and stirred continuously until all the manganese acetate was completely dissolved. Thereafter, 1.0 g fine powder of $KMnO_4$ was added slowly for 2 min then the resulting solution was further stirred for about 2 min. Afterwards, the solution was kept undisturbed for 3 days and thereafter the solution was filtered and washed with acetone repeatedly and dried at room temperature in a vacuum oven overnight and stored for further use. The vibrational spectra of functional groups of the as prepared $Mn_{12}$-ac powders and the purchased BA have been recorded using Fourier transferred infrared (FTIR) spectroscopy (Supplementary Figure S1). The variation of magnetization (M) of bulk $Mn_{12}$-ac clusters with magnetic field and temperature have been studied using Vibrating Sample Magnetometer (VSM). The temperature dependent magnetization, M(T), has been recorded to ascertain the characteristic blocking temperature of $Mn_{12}$-ac. 

\subsection{Lithographic patterning of Nb micro-tracks and Nb nano-SQUIDs}
\begin{figure}[t]
\centering
    \includegraphics[width=10cm]{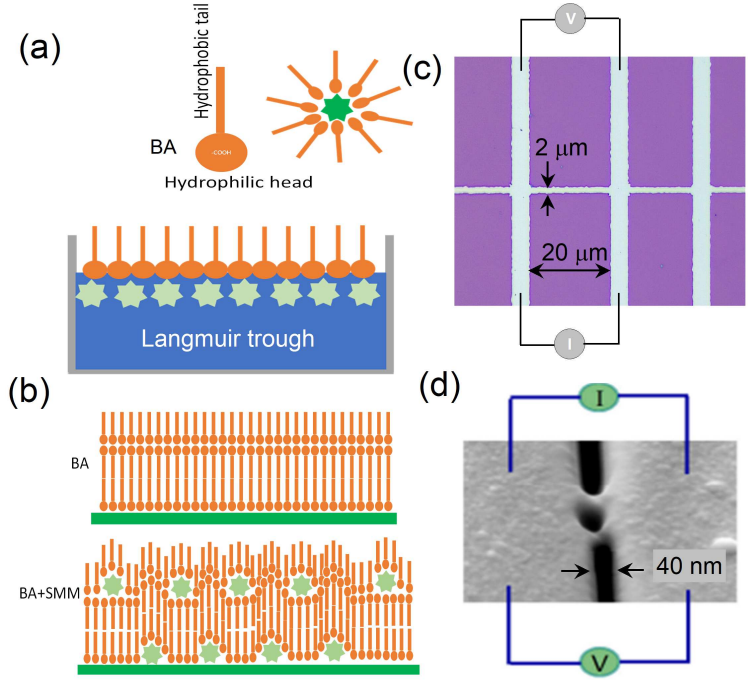}
    \caption{(a) Schematic of behenic acid (BA) molecule with hydrophobic tail and hydrophilic head, and linking of BA molecules to a single $Mn_{12}$-ac molecule in chloroform. Langmuir film of BA and $Mn_{12}$-ac mixture on the surface of water in the trough. (b) Schematic of periodic arrangement of BA layers and BA-SMM-BA layers on a hydrophilic substrate. (c) Optical image of Nb track and (d) Scanning electron micrograph of a Nb nano-SQUID.}
  \label{fgr:example}
\end{figure}
Two $\mu$m wide tracks with current and voltage contacts were patterned in on photoresist films, coated on $Si/SiO_2$ substrate, using UV photolithography technique. Thin Nb films were then deposited using DC magnetron sputtering technique at room temperature\cite{senapati2023phase} on these patterned substrates. The base pressure of the sputtering chamber was  5$\times 10^{-8}$ mBar prior to deposition of the superconducting Nb films. Sputtering of high purity Nb at a growth pressure of 1 $\times 10^{-2}$ mbar was performed while rotating the substrate under the target, for accurate control of the thickness. Subsequent to the deposition of the film, the 2 $\mu$m Nb tracks with electrodes were developed by lifting off the photoresist in acetone. An optical image of the resulting pattern is shown in the Figure 1(c). For studying the effect of the high spin magnetization of the SMM on the critical parameters of the superconducting micro-tracks, SMM layers were deposited directly on these patterned films. Prior to the deposition of the SMM layers, the superconducting critical parameters of the as grown Nb micro-tracks were measured in the temperature range of 2 K to 8 K. For detecting signatures of magnetization switching of SMM, nano-SQUIDs were fabricated on separate Nb micro-tracks using focused ion beam milling (FIB) technique. The diameter of the nano-SQUIDs were $\sim$ 40 nm. An electron microscope image of a representative nano-SQUID is shown in Fig 1(d). In this case also the magnetic field response of the nano-SQUIDs were studied prior to the deposition of the SMM layers to compare with the measurements of field response after the deposition of SMM layer.   

\subsection{Deposition of Langmuir-Blodgett film on Nb micro-tracks and Nb nano-SQUIDs}
Langmuir-Blodgett (LB) technique was used to coat the SMM in a layer by layer fashion. Hydrophilic $Mn_{12}$-ac molecules do not form a stable Langmuir film upon spreading on the water surface. Therefore, an amphiphilic matrix is required which can hold $Mn_{12}$-ac molecules on water surface. Behenic acid (BA) has been used earlier as a matrix to form stable LB films of $Mn_{12}$-ac\cite{clemente1998langmuir}. Amphiphilic BA molecule consists of a hydrophilic head and a hydrophobic tail, as shown schematically in Figure 1(a). A solution of BA and $Mn_{12}$-ac was prepared in chloroform in order to spread on the water surface. The schematic of linking of hydrophilic heads of the BA molecules to a $Mn_{12}$-ac molecule inside $CHCl_3$ is shown schematically in Figure 1(a). Upon dispersal of the solution on a Langmuir trough the hydrophobic tail of BA attached to the SMMs remains above the water surface, as shown in the Figure 1(a).
A Langmuir trough (model KSV NIMA) with two symmetrical barriers has been used to prepare the Langmuir films. The width and area of the trough was 75 mm and 24300 $mm^2$, respectively. Pt Wilhelmy plate has been used as a balance to sense the surface pressure\cite{wu2018situ} and it was kept perpendicular to the barrier. The trough and Pt plate have been cleaned with 2-propanol and Milli-Q water before use. Hydrophilic $Si/SiO_2$ and $Si/SiO_2/Nb$ substrates have been immersed inside Milli-Q Water subphase before spreading of the $Mn_{12}$-ac/BA solution. BA and $Mn_{12}$-ac with concentration of $10^{-4}M:10^{-5} M$ ratio were added to volatile $CHCl_3$ chosen as spreading agent. The prepared mixture of BA and SMM solution was homogeneously spread drop wise through out the subphase using a 50 $\mu$L micro-pipette. The system was kept undisturbed for 20 min for the evaporation of $CHCl_3$ from the surface of subphase. Then, the two symmetric barriers were compressed with a constant speed of 3 mm/min to achieve the desired surface pressure of 30 mN/m, where the $Mn_{12}$-ac/BA film was found to be in solid monolayer phase and the film was kept for 30 min before deposition where the barriers were moved to and fro with a constant of speed of 3mm/min to get a stabilized Langmuir film. After waiting for 30 min, the Langmuir film was transferred to the substrate (with in Nb micro-track or Nb nano-SQUIDs) by lifting the dipper with a speed of 1 mm/min. Further detailed parameters and procedures have been discussed in the supporting information section. The film was dried for 30 min well above the water surface. Then, the dipper was dipped again on to the subphase to deposit the second layer and lifting up to get the 3rd layer and so on.

\subsection{Electrical characterization of SMM coated superconducting Nb micro-tracks and Nb nano-SQUIDs}
In order to study the effect of magnetic flux of $Mn_{12}$-ac on a thin Nb micro-track we performed electrical transport measurements to compare superconducting parameters before and after deposition of SMM layer on the micro-tracks. All transport measurements were performed in four probe geometry. Magnetic fields were applied perpendicular to the surface of micro-strips in order to study the magnetic vortex activation process below the superconducting transition temperature. Critical current of the micro-tracks were extracted from current-voltage measurements as a function of magnetic field. The nano-SQUIDs were also connected in a four probe geometry and the characteristic SQUID oscillations were checked in bare Nb nano-SQUIDs and compared with the response of SMM coated nano-SQUIDs. SQUIDs were biased at a constant current and the voltage across the SQUID was recorded as a function of time for both types of samples in order to identify features corresponding to magnetization tunnelling in SMM.

\section{Results and discussion}

\begin{figure}[t]
\centering
\includegraphics[width=14cm]{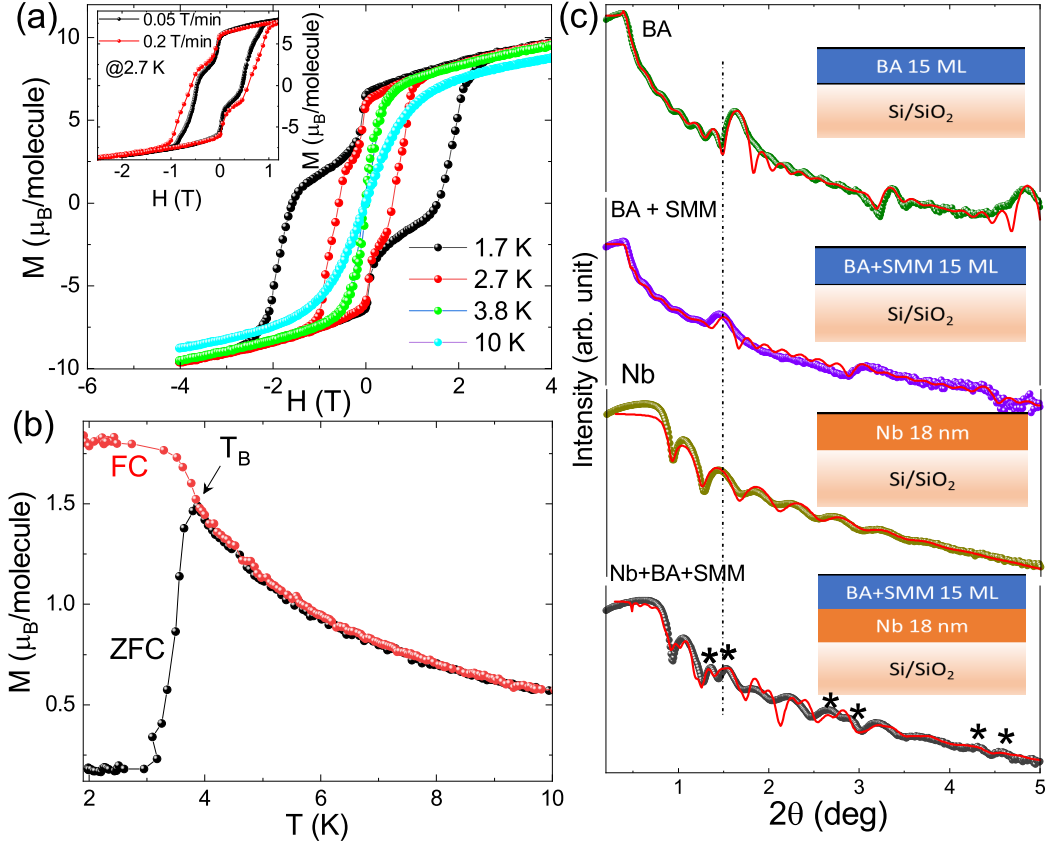}
  \caption{(a) Magnetic hysteresis loops of $Mn_{12}-ac$ powder measured at different temperatures. (b) Zero-field cooled (ZFC) and field-cooled (FC) temperature-dependent magnetization of $Mn_{12}-ac$ powder measured at 0.1 T. (c) Low angle X-ray reflectivity of 15 MLs of BA film, 15 MLs of mixture of BA and SMM film, Nb film and Nb film with 15 MLs of mixture of BA and SMM film deposited on $Si/SiO_2$. }
  \label{MHXRR}
\end{figure}

  

\subsection{Magnetic properties of $Mn_{12}$-ac powder} 
Figure 2(a) shows the magnetic field dependence of magnetization M(H) of as prepared $Mn_{12}$-ac powder measured at several temperatures close to the blocking temperature. Unlike the motion of domain walls in bulk ferromagnets, the hysteresis loop of $Mn_{12}$-ac is due to the long magnetization relaxation times\cite{sessoli1993magnetic}. To further confirm the slow relaxation behavior of $Mn_{12}$-ac, the M(H) loop has been recorded for two different sweep rates (inset of Figure 2(a)). The area of M(H) hysteresis loop and the coercive field $(H_C)$ decreases as the sweeping rate decreases from 0.2 T/min to 0.05 T/min. The steps in the hysteresis loop represents the QMT behavior of $Mn_{12}$-ac at 1.7  and 2.7 K\cite{thomas1996macroscopic}. At 1.7 K, steps in magnetization can be observed at three different fields (obtained from the dM/dH vs H curve not shown here) $\sim 0.023$ , $\sim 1.97$  and $\sim 2.3$ T. The hysteresis behavior vanishes above 4 K because thermal energy is sufficient to relax the moments in presence of magnetic field. Figure 1(b) shows the zero-field cooled (ZFC) temperature-dependent magnetization M(T) measured by applying magnetic field strength of 0.1 T. ZFC M(T) increases with decrease in the temperature, becomes maximum at $T_B\sim$ 4 K and decreases sharply with further decrease in the temperature up to 3.4 K and then, remains constant up to the lowest measurable temperature. The 0.1 T field-cooled (FC) M(T) increases with decrease in the temperature up to 3.4 K, and there is a very slow increase of M(T) with further decrease in the temperature. Large bifurcation of ZFC and FC M(T) below $T_B\sim$ 4 K is a characteristic feature of SMM signifying slow relaxation behavior of magnetization\cite{sessoli1993magnetic} in SMM molecules. In addition to the blocking temperature of 4 K, Fig 2(a) and (b) also shows that there is magnetization in the individual do not subdue above the blocking temperature, even upto 10 K, even though hysteresis vanishes above the blocking temperature(\textbf{This lines need to be modified}).  
  
\subsection{Structural characterization of $Mn_{12}$-ac multilayers deposited on Nb surface}  
 Non-destructive specular low-angle X-ray reflectivity (XRR) is an efficient technique to study the arrangement of molecules in the LB film\cite{hansen2009structure}. Figure 2(c) shows the low angle XRR of 15 layers of only behenic acid (which is the amphiphilic matrix holding the SMM in the chloroform solution) film deposited directly on $Si/SiO_2$. The diffraction pattern consists of Bragg's peaks of periodic arrangement of BA layers with one bilayer (corresponding to one full cycle of downward and upward movement of the LB dipper) as one unit cell\cite{reiche1992comparison, POMERANTZ198033} along with subsidiary peaks in between two. To simulate XRR (solid curve in Figure 2(c)), the odd 15 layers of BA has been modelled as one bottom layer and seven repeated bilayers, and the schematic of arrangement of BA layers on a hydrophilic substrate has been given in Figure 1(b). On both sides of the main peaks, the subsidiary peaks are observed because bilayers or repeated units act as diffraction gratings\cite{POMERANTZ198033}. The odd 15 monolayers (MLs) of BA consists of 7.5 unit cells, thus number of subsidiary maximum should be $7.5-2=5.5$\cite{reiche1992comparison,POMERANTZ198033}. In this case, there are 5 and 6 subsidiaries seen on the left and right of the Bragg's peak at {1.63\textdegree}. Therefore, on an average, number of subsidiary maxima is equal to 5.5. This type of XRR pattern has been observed previously in the LB film of mixture of stearic acid and 1-pyrene-dodecanoic acid\cite{reiche1992comparison}, and Mn-stearate\cite{POMERANTZ198033}. The XRR simulation yields the periodicity of about 5.4 nm for BA which is close to the reported periodicity of {22\textdegree-24\textdegree} tilted LB BA film\cite{clemente1998langmuir}. In addition, the total thickness of the film was found to be 40 nm with roughness of each layer is less than 1 nm. The XRR pattern of 15 MLs of mixture of BA and $Mn_{12}$-ac deposited on $Si/SiO_2$ is similar to 15 MLs of LB BA film deposited on $Si/SiO_2$ (Figure 1(c)). However, the Bragg's peak of periodic arrangement of BA has been shifted from {1.63\textdegree} to {1.45\textdegree} indicating intercalation of the $Mn_{12}-ac$ molecules between two BA layers, as schematically shown in the Fig 1(b). The simulated curve yields the periodicity of the two BA layers separated by a $Mn_{12}-ac$ layer to be 6.0 nm with roughness of each layer less than 1 nm and total thickness of 45 nm. The $Mn_{12}-ac$ molecule has been reported to be a disc like structure wih 1.6 nm diameter and 1.1 nm height\cite{del2011encapsulation}. The observed periodicity of two BA layer separated by a $Mn_{12}-ac$ is quite less than the sum of periodicity of BA LB film plus size of a one $Mn_{12}-ac$ molecule. This indicates that the $Mn_{12}-ac$ molecules does not form a completely (continuous) layer in between two BA layers. However, observation of no additional peak from the periodicity of BA layers indicates that SMM covers maximum part of space available in between two BA layers. The periodicity of LB film of BA and mixture of BA and SMM obtained from simulations is same as the periodicity value calculated using Bragg's equation. The thickness of Nb film determined from the XRR fitting was found to be 18 nm (Figure 2(c)). In case of Nb with SMM, all subsidiary peaks of the repeated bilayers were not visible due to the dominance of diffraction intensity of Nb film over BA and SMM. However, the Bragg's diffraction peaks of periodic arrangements have do show up in the diffraction  pattern marked with asterisk symbols on the Fig 2(c) at different $2\theta$ values. The periodicity and thickness of LB film of mixture of BA and $Mn_{12}-ac$ on $Si/SiO_2/Nb$ was same as the LB film grown on $Si/SiO_2$. The Bragg's peak of periodicity of LB film on Nb appeared at almost same $2\theta$ value of LB film on $Si/SiO_2$ which is indicated by the dashed vertical line in Figure 2(C). The above discussion shows that the SMM layers formed by the LB technique does not produce continuous layers films of SMM, possibly due to the fact that each $Mn_{12}$-ac molecule attaches to several behenic acid molecules in the chloroform solution, as shown the schematic Fig 1(a). When dispersed on the water surface in the Langmuir trough the behenic acid molecules open up to allow hydrophobic ends to stay above the water surface. As a result the separation between the individual SMM molecules naturally increases, forbidding a continuous film of SMM in this method. This is however, an ideal structure for the studies of the effects of SMM layers on superconducting films and devices as dipolar interaction between the SMMs can be neglected.

\subsection{Superconducting properties of $Mn_{12}$-ac coated Nb micro-tracks} 
Figure 3(a) shows the comparison of temperature dependent resistance (R(T)) of a bare Nb track and  Nb track with 21 layers of LB $Mn_{12}$-ac film in zero field. The $T_C$ of 18 nm bare Nb micro-track was $\sim 6$ K in our case. As the thickness of the Nb film was less than the coherence length of Nb\cite{duan2021normal}, one expects a decrease in the transition temperature compared to the bulk value of 9 K for Nb. Fig 3(a) shows that the proximity of $Mn_{12}$-ac SMM layer decreases the $T_C$ of Nb track to some extent. This is due to the breaking of cooper pairs by the local magnetic moment of quasi-isotropic distributed $Mn_{12}$-ac. We mention here that the data in Fig 3 is from the same track which was measured before and after the deposition of SMM. To confirm further the reduction in $T_C$, LB films of mixture of BA and $Mn_{12}$-ac of different layers have been grown on Nb track successively and compared in the supplementary Figure S3. The $T_C$ of bare Nb decreases with increase in number of layers signifying gradual increase in the strength of proximity effect of the SMM layers with increasing thickness. The change in resistance of Nb track with and without SMM predominantly near the superconducting to normal metal transition region. However, as expected, there is no change in the resistance in the superconducting region due to dominance of conduction of Nb. The insets of Figure 3(a) show the R(T) of a Nb micro-track in the presence of different magnetic fields applied perpendicular to the film plane, before and after coating with $Mn_{12}$-ac SMM. In type-II superconductors, the superconducting state to normal state transition is described by the thermally activated flux flow (TAFF) model\cite{TAFF}. The TAFF model predicts that vortex motion across Arrehenius type pinning barrier causes the change in resistance in the transition regime. The change in resistance is given by $R(T,H)=R_0e^{-U/k_BT}$, where $R_0$ is the pre-exponential factor, U is the vortex activation energy and $k_B$ is the Boltzmann constant\cite{TAFF, PhysRevB.103.L180503, ciccullo2018interfacing, pratap2023optimization}. The temperature and field dependence of U is given by $U=U_0(H)(1-T/T_c)$\cite{PhysRevB.103.L180503}. Magnetic flux emanating from the $Mn_{12}$-ac molecules may penetrate the superconducting layer and consequently may change the vortex activation process. Therefore, change in the activation energy may be used as a measure of the effect of the SMM proximity layer. In the insets of Figure 3(b) we plot the Arrhenius plots ($lnR$ vs 1/T) for different magnetic fields where $lnR$ varies linearly with 1/T in the transition region shown by the red lines. $U$ of Nb tracks with and without SMM has been extracted from the slopes of the linear fit to the Arrhenius plots in Figure 3(b). $U$ of bare Nb film and Nb film with SMM decreases with increase in the magnetic field strengths. The main panel in Fig 3(b) shows that $U$ of SMM coated Nb track is clearly higher than the Nb film which indicates that the network of $Mn_{12}$-ac molecules in the SMM layer helps in enhancing the pinning effect of the superconducting track. 
\begin{figure}[t]
\centering
\includegraphics[width=12 cm]{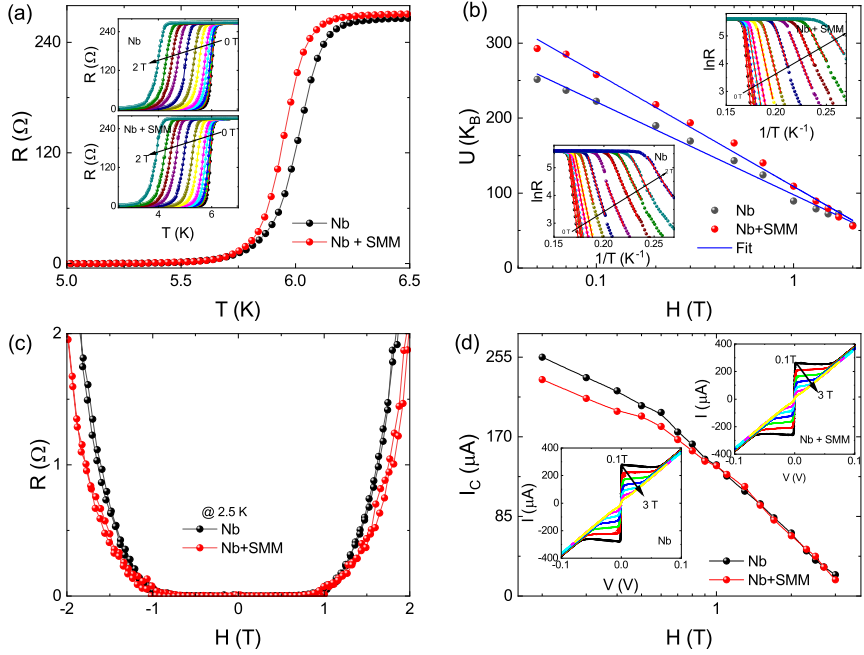}
  \caption{(a) Temperature-dependent resistance R(T) at 0 T and at different magnetic fields (see insets) of bare Nb film and Nb film with SMM. (b) Field dependence of $U$ of bare Nb and Nb with SMM and the red lines are the fits to the eq. $U\propto lnH$. The insets show the Arrhenius plots of resistance $lnR$ vs 1/T at different magnetic fields and the red lines are the linear fits. (c) Variation of resistance with magnetic field R(H) at 2.5 K of Nb and Nb with SMM. (d) Field dependence of critical current $(I_C)$ of Nb and Nb with SMM at 2.5 K. The insets show the I-V characteristics of Nb and Nb with SMM at 2.5 K.}
  \label{fgr:example}
\end{figure}
Interestingly, the difference in the $U$ values obtained for Nb film with and without SMM decreases with increasing magnetic field strength. In fact, activation energies in the two cases almost match at the a field slightly higher than 1 T, which can be identified as the saturation field for the SMM particles from Fig 2(a). At around 1 T the magnetic moment of all the SMM particles aligns with the magnetic field and, concurrently, the additional vortex unbinding energy due to the SMM layer vanishes. This effect clearly indicates that the random orientation the moments in the SMM layer promotes vortex-antivortex binding. The activation energy $U$ varies with magnetic field following the empirical relation $U=U_{0} ln(H_0/H)$ where $U_{0}$ is the characteristic vortex unbinding energy and $H_0$ $\simeq$ $H_{C_2}$ for clean superconductor \cite{banerjee2019restoring}. The blue solid lines shown in Fig 3(b) are the fits to the data following this model. This type of logarithmic variation of $U$ indicates the 2D nature of vortices in the presence magnetic field perpendicular to the film plane\cite{PhysRevB.103.L180503}. The fit yields $U_{0}$ of bare Nb track as $53.72$ K which is less than the $U_{0}=65.5$ K of Nb with SMM. 

Apart from the vortex-antivortex unbinding, the vortex pinning effect in the superconducting state can also be a marker for studying the effect of SMM layer on the superconducting layer. For this purpose, we measured the magnetic field dependence of resistance of bare Nb track and SMM coated Nb track at $T=2.5$ K have been compared in Figure 3(c). The resistance of Nb track with and without SMM remains same in the superconducting state up to 1 T, however, the resistance starts increasing with further increase in the magnetic field strength. At relatively higher magnetic fields, the resistance of SMM coated Nb track appears lower than the bare Nb track which is caused by an enhanced vortex pinning induced by the $Mn_{12}-ac$. The insets of Figure 3(d) show the current-voltage (I-V) characteristics of bare Nb and Nb with SMM measured at 2.5 K in the presence of different magnetic fields applied perpendicular to the film plane. A small decrease in the critical current $(I_C)$ of Nb because of the $Mn_{12}$-ac layer is noticed at lower fields. However, at higher fields these two curves coincide. This can only be explained if there is an effective enhancement in the pinning effects.

In the previous section we discussed rather indirect, though relatively simple, measurements to sense the flux effects of SMM layers using superconducting films. These measurements inherently provide an average response of a collection of SMM particles. Quantum activity at the level of individual molecules is inaccessible to these measurements. In this section we attempt to observe some signature of flux activity at the molecular level. In order to search for such signatures, the only possible method is to use SQUIDs which can convert magnetic flux changes to a measurable voltage in a very sensitive manner. In order to restrict the amount of SMM particles exposed to the SQUID loop we have fabricated constriction-based Nb nano-SQUIDs using Ga focused ion beam (FIB). The dimensions of both the weak links were kept close to the coherence length of Nb which also avoided any hysteretic behaviour in the I-V characteristics \cite{troeman2007nanosquids}. 21 MLs of LB $Mn_{12}-ac$ film was deposited on the SQUID to detect the MQT behaviour. The scanning electron microscope (SEM) micrograph of the nano-SQUID with a loop diameter of 60 nm on a 40 nm thick Nb film is shown in the inset of Figure 4(a). The loop inductance was found to be 0.014 pH. For attaining enhanced voltage response and high critical current density (0.025 A/$\mu m^2$), we keep both the junctions' size very small to be $20\times50$ $(w\times l)$ $nm^2$. The smaller junction may bring kinetic inductance of the loop which can dominate over loop inductance \cite{jose2017nanosquids}. By considering ($\lambda$ = 120 nm), the kinetic inductance for each junction is found to be ($L_{KE} = \mu_0 \lambda^2 l /wt$) 1.13pH \cite{luomahaara2014kinetic, van1981principles}. Fabricated nano SQUID device have $\beta_L \ll 1$, and the constricted Nb within coherence limit provides the necessary condition for non-hysteretic I-V measurements \cite{voss1980niobium, granata2011noise}.  In Figure 4(a) we compare the R(T) of the fabricated nano-SQUID device with and without SMM thin films. The resistance was measured at 1$\mu$A current while warming from the lowest measurable temperature. The $T_C$ of the nano-SQUID remains unaffected (7.7 K) after the deposition of 21 layers of LB $Mn_{12}-ac$ film. This indicates that the magnetic moment of SMM is practically ineffective for reducing the $T_c$ of 45 nm thick Nb film, unlike the case in the previous section, where the Nb film was only 18 nm thick.

To check the characteristics signature of a SQUID device, we have measured the voltage across the device (V)  as a function of magnetic field (H), at the lowest possible temperature (2 K). The comparison of the SQUID response without the SMM and with the SMM is shown in Figure 4(b). We must mention here that these measurements were performed with magnetic field perpendicular to the surface of the SQUID loop. A bias current of 100 nA was used in this case. Figure 4(b) shows the periodic oscillations in the voltage across the SQUID which are rather triangular in shape as expected in the short Josephson limit \cite{hasselbach2002micro}. Although the critical current for this SQUID is 40 $\mu$A, we have chosen a low current for the SQUID operation. The oscillation period was found to be $\sim$400 Oe at low fields which increased slightly at the higher fields, possibly due to the increase in vortex count \cite{ku2016decoherence}. The flux sensitivity for this nano-SQUID was found to be $\sim$1.5 mV per flux quantum (mV/$\phi_0$). 
We do not observe any difference in the oscillation period in the Fig 4(b) after the SMM layer deposition in perpendicular field orientation. 
\subsection{Signature of magnetization tunnelling in $Mn_{12}$-ac coated Nb nano-SQUID}
\begin{figure}[t]
\centering
\includegraphics[width=14 cm]{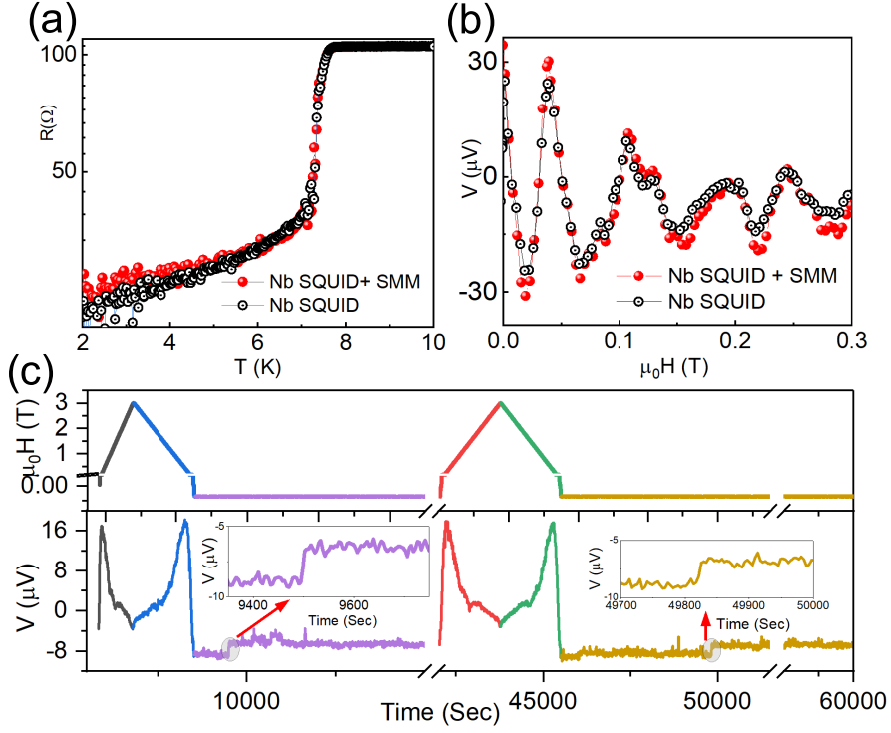}
\vspace{-1cm}
  \caption{Detection of Microscopic tunnelling in SMM thin films using nano-SQUID: (a) RT plot of a Nb nano-SQUID with and without SMM thin films. False colour SEM image of the fabricated device shown in the inset, with a scale bar of 50 nm. (b) The same device shows no change in SQUID oscillations in the VH plot for the perpendicular magnetic field at 2 K. (c) The voltage response of the SQUID device with SMM thin film at $-0.02 $ T was recorded after coming from a high field for 8 hours. The inset shows voltage response in a parallel magnetic field coming from 3 T to $-0.02$ T for the same device.}
  \label{fgr:example}
\end{figure}

The non-hysteretic SQUID here can be used as a voltage-to-flux converter, which we intend to employ in this case to detect local magnetic flux change arising from magnetization tunnelling in $Mn_{12}$-ac. A high magnetic field can magnetize the SMM along the field direction, which can then relax to the ground state by tunnelling across the high anisotropy barrier. This process can be either purely quantum mechanical tunnelling of magnetization or thermally assisted tunnelling at temperatures close to the blocking temperature. In both cases, the process is completely random in time and results in a change in magnetization. A nano-SQUID placed close to the SMM particles can sense the change i the magnetic flux due to the tunnelling of magnetization from one state to another. Since it is a random process, the SQUID voltage has to be monitored over time. For the SMM coated nano-SQUID we performed the experiment in presence of a parallel magnetic field. To start with, a high magnetic field of +3 T was applied in parallel configuration to saturate the moment of SMM in the field direction. In the parallel field configuration, the external magnetic flux passing through the nanoSQUID loop is ideally zero, and therefore, it is rather insensitivity to parallel external magnetic fields \cite{wernsdorfer1996nucleation, vohralik2009nanosquid}. Subsequently, the magnetic field was ramped down to a small negative magnetic field of $-200$ Oe. The SQUID voltage was observed at this point for a long time at a fixed bias. Fig 4(c) shows the results of this experiment. The top panel of Fig 4(c) shows the magnetic field ramping protocol and the bottom panel plots the voltage across the SQUID device coated with a 21-layer SMM film. While ramping the magnetic field, an oscillating voltage across the SQUID was observed, even though the field was applied parallel to the SQUID loop. This is due to some finite angle between the field and the SQUID loop in the experimental setup, as apparent from the very high period of oscillation in this range. We observed a sudden jump in the voltage during the time frame where the external magnetic field was kept constant. On repeating the same experiment we observed a similar voltage jump at a different time. 
In contrast, when the same experimental conditions were applied to the bare Nb nano-SQUID, no analogous voltage jumps were observed over a very long waiting period. The observed jumps in the SQUID voltage can only be explained by a change in magnetic flux through the SQUID, only possible via magnetization tunnelling phenomenon of SMMs. The voltage response we observed was reaffirmed when another SQUID of comparable size on the same chip yielded identical results. Therefore, even though the nature of magnetization tunnelling can not be ascertained in this experiment, we show that a nano-SQUID in close proximity of SMMs can detect magnetization tunnelling phenomenon. 

\subsection{Conclusions}
In conclusion, we have shown that superconducting systems can effectively sense the magnetic flux of magnetic molecules when placed in close contact. Transport measurements were performed on $Mn_{12}$-ac coated thin superconducting Nb stripes near the superconducting transition and also deep inside the superconducting temperature regime. The Langmuir-Blodgett technique allowed us to form layered SMM films, albeit with low density of SMM. By comparing the critical current of Nb microstrips before and after SMM coating We showed an increase in the vortex pinning effect. In the superconducting transition regime, an enhanced vortex-antivortex unbinding energy was found in the low field regime. In the saturation field regime, the SMM layer did not have an impact on the superconducting transition. Nb-based nano-SQUIDs with very small active loops were fabricated using the FIB technique in order to detect signatures of local change in magnetization in the $Mn_{12}$-ac molecules arising from quantum tunnelling of magnetization. By recording the SQUID voltage as a function of time, we found random jumps in the voltage across the SQUID, which can only be explained by a sudden change in the magnetization of SMM in the vicinity of the SQUID loop. No such jumps were observed in the SQUIDs without the SMM layer. Therefore, the sudden jumps in the SQUID voltage are very likely to be signatures of magnetization tunnelling. More work on SMM-coated nano-SQUIDs is needed to quantify the nature of the magnetization tunnelling signatures observed in this method. This work shows the feasibility of hybrid spintronics devices combining superconductivity with the wide variety of available high-spin molecular magnets.    


\begin{acknowledgement}

The authors thank the National Institute of Science Education and Research (NISER), Department of Atomic Energy, Government of India, for funding the research work through project number RIN-4001 and RIN-4002. 

\end{acknowledgement}



\newpage
\section{\centering Supplementary information}

\subsection{Deposition of Langmuir-Blodgett films of SMM}
\begin{figure}[b]
\centering
\includegraphics[width=12 cm]{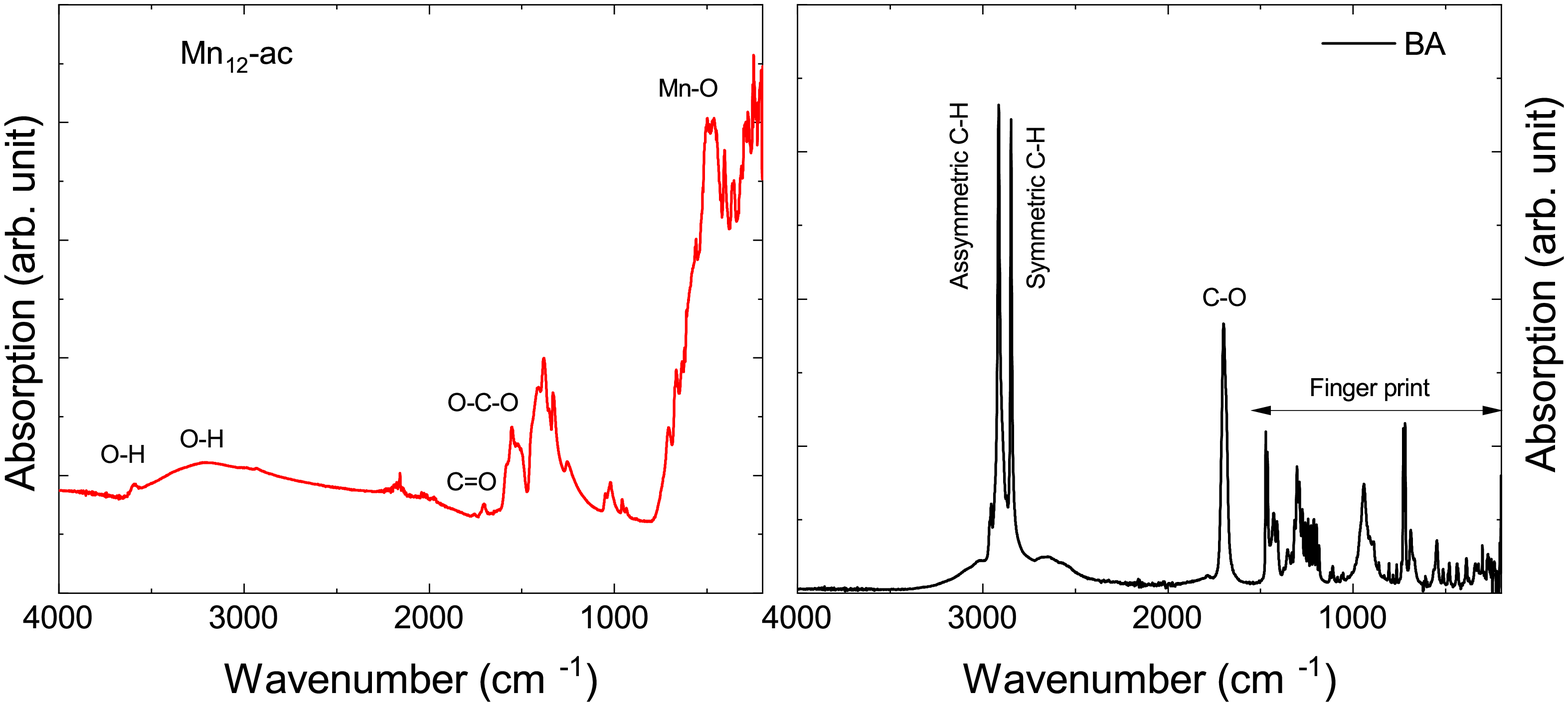}
  \caption*{ Figure S1: Fourier transformed infrared spectroscopy spectra of $Mn_{12}$-ac powder and behenic acid.}
  \label{}
\end{figure}
Before making of solution of mixture of BA and $Mn_{12}$-ac, the characteristic vibrational peaks of functional groups of as prepared $Mn_{12}$-ac and the purchased BA have been checked using FTIR spectroscopy. Figure S1(a) and S1(b) shows the FTIR spectra of $Mn_{12}$-ac and purchased BA, respectively and the different vibrations modes have been identified in the figure according to the previous report\cite{VERMA201776, B819582J}. Figure S2(a) shows the complete isotherm of BA floats on the subphase where the variation of surface pressure (SP) with mean molecular area (Mma) has been shown. The two-dimensional analogue of a pressure is called SP, $SP= \gamma - \gamma_0$ where $\gamma$ is the surface tension of subphase in the absence of monolayer and $\gamma_0$ is the surface tension in the presence of monolayer. Distinctive regions of the isotherm curve represent different phases of BA formed during the compression of the two symmetric barriers (Figure S2(a)). The SP remains close to zero as the Mma decreases up to 26 $\AA^2$ which represents the gaseous state of BA. In further decrease in the Mma by compressing the barriers, the SP increases with decrease of Mma represents the liquid state of BA. There is a clear transition from liquid to solid phase, the SP increases sharply with the compression of the barriers where the BA acid is in the form of solid monolayer. The Mma in the solid phase of BA is found to be around 21 $\AA^2$ by extrapolating the linear SP line to abscissa. This value is close to the reported value of BA \cite{rehman2016morphology, zhavnerko2002composite}. The observed isotherm is quite similar to the previously reported comparable systems \cite{gorbachev2022langmuir}. The isotherm of mixture of BA and $Mn_{12}$-ac in $10:1$ M ratio is significantly different to the isotherm of only BA (Figure S2(a)). The liquid state to solid phase transition is not sharp in this case, which clearly indicates the $Mn_{12}$-ac has been attached at the hydrophilic head of the BA matrix in addition to the less Mma in the solid monolayer state (Figure S2(a)). The solid monolayer of mixture of BA and $Mn_{12}$-ac have been transferred to substrate at target $SP=30$ mN/m\cite{clemente1998langmuir}.

\begin{figure}[t]
\centering
\includegraphics[width=12 cm]{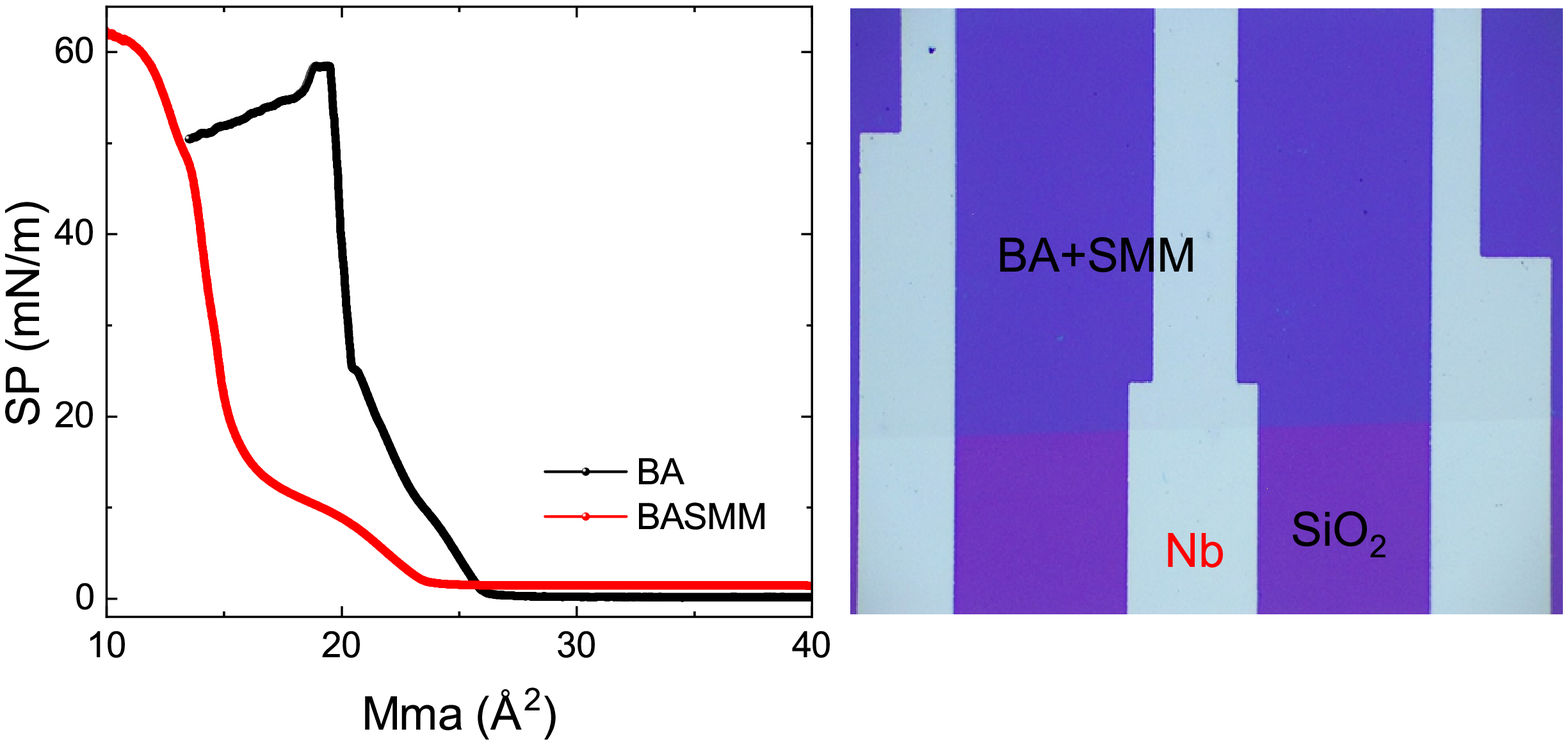}
  \caption*{Figure S2: Complete isotherm of (a) BA and mixture of BA and $Mn_{12}$-ac SMM in the 10:1 M ratio. (b) Optical image of LB film of mixture of BA and $Mn_{12}$-ac SMM grown on Nb pattern.}
  \label{}
\end{figure}

\newpage
\subsection{Superconducting transition in Nb micro-tracks with various SMM layer thickness}
In order to verify that the SMM layers affect the transition temperature of the Nb micro-track we have measured the same Nb track after 3, 9, 15, and 21 layers of LB film of BA and SMM. A systematic change in transition temperature is observed in the supplementary Figure S3. 
\begin{figure}
\centering
\includegraphics[width=10cm]{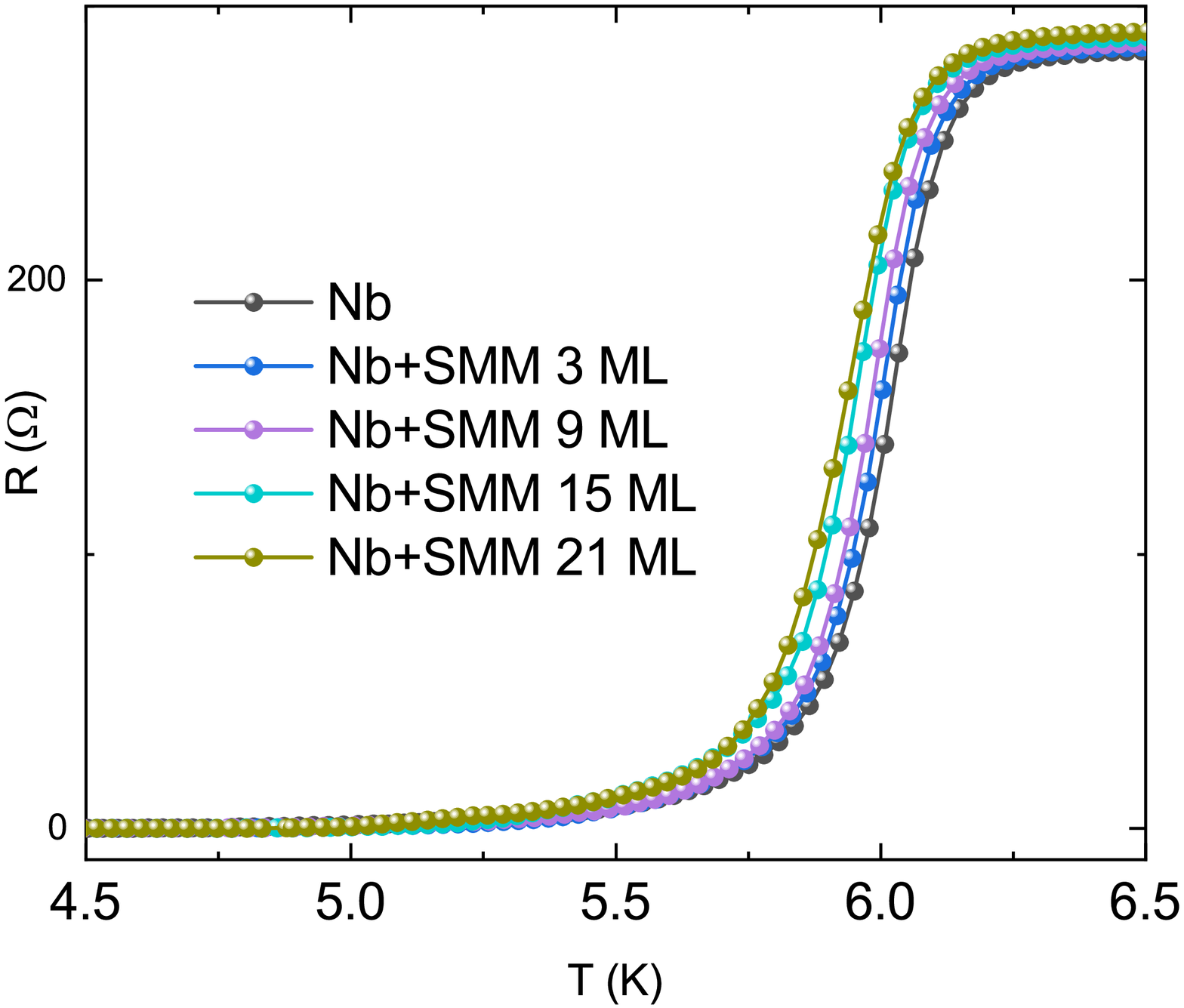}
  \caption*{Figure S3: Temperature dependence of resistance of bare Nb track and Nb track with different layers of mixture of BA and SMM.}
  \label{}
\end{figure}

\bibliography{SMMSQUID}
\end{document}